# Microscopic Mechanism of Stereochemically Active Lone Pair Studied from Orbital Selective External Potential Calculation

Yongping Du,[1] Hang-Chen Ding,[2] Li Sheng,[1] Sergey Y. Savrasov,[3] Xiangang Wan,[1,*] and Chun-Gang Duan[2,4,*]

[1]*Department of Physics and National Laboratory of Solid State Microstructures, Nanjing University, Nanjing 210093, China*
[2]*Key Laboratory of Polar Materials and Devices, Ministry of Education, East China Normal University, Shanghai 200062, China*
[3]*Department of Physics, University of California, Davis, One Shields Avenue, Davis, CA 95616, USA*
[4]*National Laboratory for Infrared Physics, Chinese Academy of Sciences, Shanghai 200083, China*

The nature of the stereochemically active lone pair has long been in debate. Here, by application of our recently developed orbital selective external potential (OSEP) method, we have studied the microscopic mechanism of stereochemically active lone pairs in various compounds. The OSEP method allows us to shift the energy level of specific atomic orbital, therefore is helpful to identify unambiguously the role of this orbital to the chemical and physical properties of the system we are interested in. Our numerical results, with compelling proofs, demonstrate that the on-site mixing of cation valence $s$ orbital with the nominally empty $p$ orbitals of the same subshell is crucial to the formation of lone pair, whereas the anion p orbital has only small effect. Our detailed investigation of Sn and Pb monochalcogenides show that structures of these systems have significant effects on lone pairs. In return, the formation of lone pair, which can be controlled by our OSEP method, could result in structural instabilities of Sn and Pb monochalcogenides.

PACS numbers: 74.20.Pq, 74.70.-b

The cations with a formal $ns^2np^0$ electronic configuration usually display novel properties, and it is widely believe that the so called $ns^2$ electronic lone pair is responsible for its stereochemically activity[1-4]. The stereochemical activity of lone pair electrons has been shown to cause Jahn-Teller geometry distortions [5, 6], large optical response in $TeO_2$-based materials [7], as well as the ferroelectricity in $BiFeO_3$ [8] and $BiMnO_3$ [9]. Lone pair has also been used to explain the anisotropies of thermal expansion [10], piezoelectric and elastic properties[11] and optoelectronic properties [12]. Thus the stereochemical activity of lone pair had been studied extensively over the past seven decades in both chemistry and physical community [1-4, 8, 9, 13-28].

However, despite the tremendous efforts that have been devoted [1-4, 8, 9, 13-28], the microscopic understanding of the lone pair is still not fully established. In the early valence-bond stereochemistry theory, Sidgwick and Powell [1] proposed that every electron pair has the same electrostatic repulsion regardless it is lone pair or bonded electron pair. This theory is, however, too simple to explain some of the more quantitative aspects of molecular geometry. Later Gillespie and Nyholm [2] suggested that lone pair has a larger repulsion with others than the bonded electron pair. This theory successfully explained the shape of many small molecules such as NH3, $H_2O$, $SF_4$, $SnCl_2$, *etc*. Orgel [3] explained lone pair effects by invoking intra-atomic mixing of cation $s$ and cation $p$ orbitals. Since the $s$ and $p$ orbitals have different parities, thus cannot hybridize in the environment with inversion symmetry. Nevertheless, if the energy gain due to the *s-p* hybridization induced by broking the inversion symmetry is over the energy cost to excite the electron from $s$ to $p$, the system has to undergo a distortion to break the symmetry and make the electron's distribution asymmetric. This theory naturally explained the structural distortion and the large dielectric response of the system with lone pair [3]. In this theory the relative energy of nominally filled $ns$ and nominally empty $np$ states of the cation is critical to the formation of stereochemically active lone pair, thus it seems hard to understand why the formation of lone pair also depends on the anion [4]. For example, the tetragonal distortion of the cubic rocksalt structure takes place only for PbO with asymmetric electronic distributions around $Pb^{2+}$ ion, whereas PbS remains undistorted in a rocksalt phase with symmetric electron distribution around $Pb^{2+}$ ion [4].

In additional to phenomenological arguments, the mechanism of lone pair had also been studied extensively by density functional calculation [4,13-28]. Watson [13, 14] showed that contrary to the traditional picture, the 6$s$ electrons in PbO are certainly not chemically inert. Their numerical results clearly showed that Pb 6$s$ and O 2$p$ bands have considerable hybridization [13, 14]. The hybridization between cation $ns$ lone pair and anion $p$ band had also been claimed by electron localization function (ELF) [8, 9] study as well as cooperative orbital Hamiltonian overlap method [8, 9]. On the experimental side, the importance of hybridization had also been demonstrated by the high-resolution X-ray photoemission and soft X-ray emission spectroscopies [27], with findings that a high density of metal 6$s$ states is located at the bottom of the valence band instead of close to the Fermi level. These results call into question about the long-established conventional explanation of the formation of the $Pb^{2+}$ lone pair [3]. Based on the numerical results, Watson and his colleague asserted that the mixing of nominally empty Pb 6$p$ state with the filled antibonding O 2$p$ and Pb 6$s$ states causes the formation of lone pair [4, 13, 14, 17, 20, 23, 27]. There is also numerical work emphasizing only the cation $s$ and anion $p$ state, without involving the cation $p$ states [28].

In this article, we apply orbital selective external potential (OSEP) method, which is previously called constrain orbital-hybridization method [29], to investigate the origin of lone pair and its stereochemically effect. The OSEP method enables us to apply an external field to shift the energy level of specific atom orbital, consequently tune the atomic orbitals which join the hybridization, and clearly clarify the contribution from various orbital. Firstly, we applied this method on α-PbO, with respective shifting of Pb 6*s* and 6*p* subshells, O 2*s* and 2*p* subshells, and found that the O 2*s* and 2*p* subshell have little effect on lone pair, while Pb 6*s* and 6*p* can change the shape of lone pair. Then we investigate $PbTiO_3$, $Bi_2O_3$, $BiFeO_3$, we found out the same results as the α-PbO. At last we perform a detailed investigation of SnX (X=O, S, Se, Te) and PbX (X=O, S). Our numerical results give direct and clear evidence about the effect of lone pair on lattice structure.

We perform the electronic band-structure calculations within the local density approximation (LDA) frame using the full-potential linearized-muffin-tin-orbital LMTO method [30]. We also cross-check the results by the commonly used Vienna *Ab-Initio* Simulation Package (VASP) [31]. The consistency of our results for two sets of calculations is satisfactory. For consistency, the results presented in this paper are obtained by LMTO, unless otherwise specified.

We have introduced the OSEP method in previous study [29]. Here, to extend its applications, we would like to rephrase this method as follows.

The spirit of the OSEP approach is to introduce a special external potential. Different from the realistic external potential, this potential is orbital sensitive, i.e., only certain appointed orbital can feel it. Though this orbital sensitive potential is originally proposed for theoretical purposes, there is possibility such potential could exist in nature. Specifically, we define a projector operator $|inlm\sigma\rangle\langle inlm\sigma|$, which only allow the external potential $V_{ext}$ influence the specific atomic orbit $|inlm\sigma\rangle$. Here *i* denotes the atomic site, and *n, l, m, σ* are the main quantum number, orbital quantum number, magnetic quantum number, spin index, respectively.

The new Hamiltonian can be written as:
$$H^{OSEP} = H_{KS}^0 + |inlm\sigma\rangle\langle inlm\sigma|V_{ext},$$
where $H_{KS}^0$ is the original Kohn-Sham Hamiltonian which includes all the orbital-independent potential. Using this new Hamiltonian, we can solve the new secular equation in the framework of density functional theory (DFT), in a self-consistent way. In our scheme multiple orbital-dependent potentials can be applied to the system simultaneously, providing great flexibility to study various effects to the physical or chemical properties of the system. We should point out that our OSEP method is analogous to the LSDA+*U* method [32], in the sense that the applied potential is orbital-dependent. However, due to the different physical meaning of these potentials, the OSEP method may also be applied in the systems other than strongly correlated. Indeed, our strategy allows the co-application of LSDA+*U* and OSEP methods.

As the strength of hybridization between two orbits is strongly dependent on their energy difference, by applying external field to shift the energy levels of these orbitals, we can effectively weaken (or strengthen) the hybridization. This is the main motivation we developed the OSEP method, though its application is not limited for this purpose. We have applied this technique to study perovskite ruthenates [29] and europium monochalcogenides [33], and the results are successful. Here we use this method to study the mechanism of lone pair.

As a typical lone pair active compound, α-PbO had been studied extensively [4, 13, 14, 16, 23, 27, 34-38]. We therefore chose α-PbO to initialize our study. We first briefly introduced the structure and electronic structure of α-PbO, and demonstrated graphically the concept of lone pair. Then we applied the OSEP method to detail possible causes of the formation of such lone pair.

Pb cation in α-PbO has a formal $6s^26p^0$ electronic configuration. If the $6s^2$ electron pair is chemically inactive, the charge distribution around this cation should be spherical, and the compound should have inversion symmetry. However, due to the hybridization with the nominally empty 6*p* either directly [3] or indirectly through O-2*p* orbital [4, 13, 14, 23, 27], the lone pair loses its spherical symmetry and is projected out one side of the cation, resulting in asymmetry in the Pb coordination and distorted crystal structure [13, 14]. α-PbO crystallizes in a tetragonal structure with space group *P*4/*nmm*, and the Pb and O atoms occupy the 2*c* (1/4, 1/4 ,*z*) and 2*a* (3/4, 1/4, 0) sites, respectively [39]. Our optimized lattice parameters (*a*=3.95, and *c*=4.87 Å) are close to the experimental data (*a*=3.96 Å, and *c*=5.01 Å) [39]. The theoretical internal atomic coordinate (*z*=0.24) is also in good agreement with the experimental result (*z*=0.26) [39]. Based on the obtained lattice structure, we performed total-energy calculation for α-PbO. As shown in the plot of partial density of states (PDOS) of α-PbO (shaded plots in Fig. 1), O-2*s* state locating around -17 eV has negligible interaction with other orbitals. Pb-6*p* bands distribute mainly above the Fermi level, but due to the strong interactions with O-2*p*, Pb-6*p* also has considerable distribution between -6.0 and 0.0 eV, where the O-2*p* states dominate. Pb-6*s* states appear mainly at -10.0 to -6.0 eV. There is considerable mixing between Pb-6*s* and O-2*p* states, as shown in Fig. 1, which confirms the strong hybridization between Pb-6*s* and O-2*p* states, as suggested by previous theoretical studies [4, 13-15].

As mentioned earlier, the strength of hybridization between two orbitals strongly depends on their energy difference, therefore the lone pair should be sensitive to the shifting of special orbital if this orbital participate in the formation of lone pair. We thus perform the OSEP approach to analyze the mechanism directly. Note that the OSEP method allows us to gradually shift the atomic orbital, therefore we can see a continuous change of the band structure caused by certain band shifting. Since the results for different field is qualitatively similar, we only show the results with external field to shift the on-site orbital energy by 6 eV. To show the effectiveness of our method, we first apply field to downshift O-2*s* band. The PDOS after downshifting O-2*s* band by 6 eV is shown in Fig. 1(a). For comparison, we also draw the original PDOS as a shaded plot. Due to the fact that O-2*s* almost has no hybridization with other orbits, the down-shift of O-2*s* states has little influence on the O-2*p* and Pb states, as shown in Fig. 1(a). We note that the down-shift to deeper energy level results in a narrower and sharper PDOS of O-2*s* state. This demonstrates that our method is not merely a rigid shift of energy states, and the results are self-consistent instead.

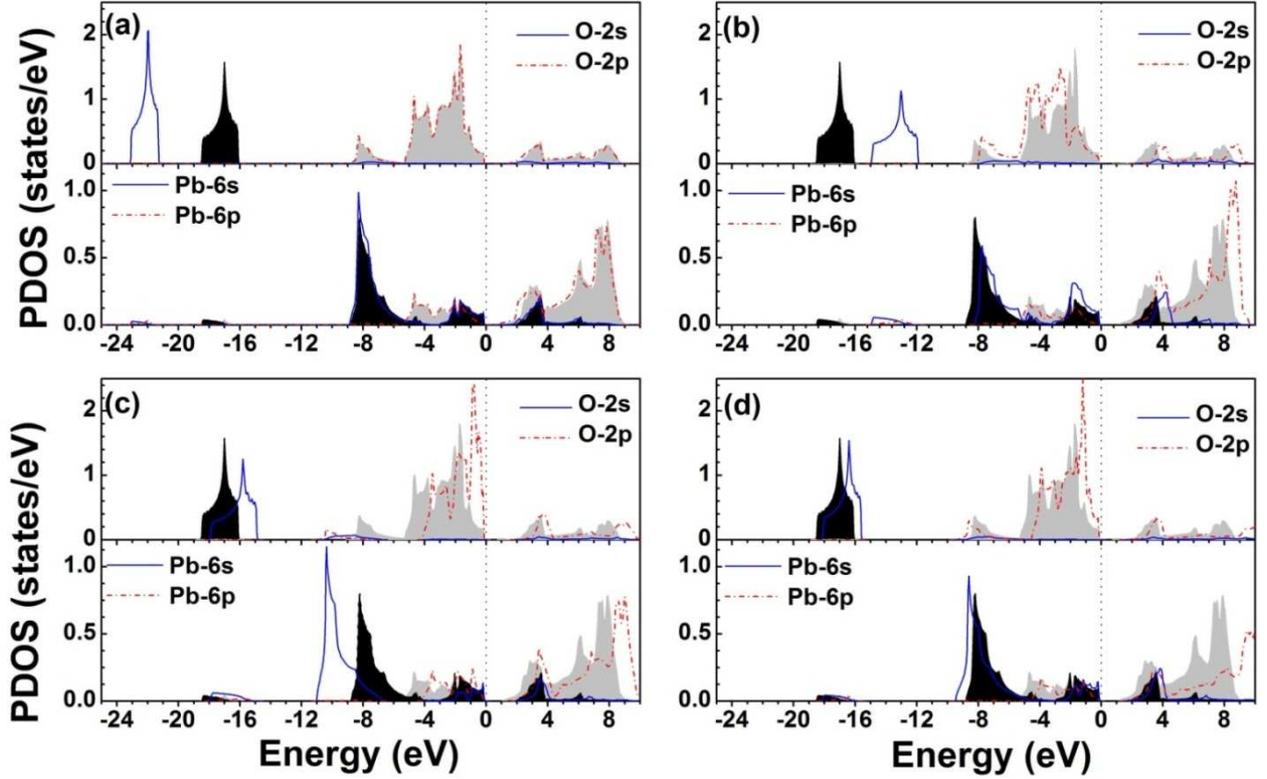

FIG. 1 (color online). Partial density of states (PDOS) of α-PbO with: (a) shifting down O-2$s$; (b) shifting down O-2$p$; (c) shifting down Pb-6$s$; (d) shifting up Pb-6$p$. O-2$s$ and Pb-6$s$ are shown as blue straight lines, O-2$p$ and Pb-6$p$ are shown as red dashed dotted lines. For comparison, PDOS of α-PbO with no applying field are shown as shaded plots: black for O-2$s$ and Pb-6$s$ and light gray for O-2$p$ and Pb-6$p$. The Fermi level is set to be at zero, shown as vertical dotted lines.

We also apply field to downshift O-2$p$ states. As the top valence bands are dominated by O-2$p$ states, downshift them will also bring down the Fermi energy, therefore the deep energy states such as O-2$s$ state is equivalently shifted up as shown in Fig. 1(b). The Pb-6$s$ and Pb-6$p$ states are also affected by the down-shift of O-2$p$ states, due to the strong hybridization of O-2$p$ with Pb-6$s$(6$p$) states. This results in weaker hybridization of O-2$p$ with Pb-6$p$ states yet stronger hybridization of O-2$p$ with Pb-6$s$ states, as evidenced by the fact that the Pb-6$p$ bands becomes considerably sharper whereas the peak of Pb-6$s$ state decreases [Fig. 1(b)]. We note that, also due to the strong hybridization of O-2$p$ with Pb-6$s$(6$p$) states, the actual shifting energy of O-2$p$ is not exactly as we applied. This is different from the case of O-2$s$, where due to the chemical inertness of this orbital, the actual shifting energy as shown in Fig. 1(a) is almost the same as we applied.

As can be expected, shifting down the Pb-6$s$ band decreases the strength of hybridization between Pb-6$s$ and O-2$p$ states. This is confirmed by results shown in Fig. 1(c). We notice that O-2$p$ bands become considerably narrower and their peaks become higher. Down-shifting Pb-6$s$ band also decreases the energy difference between O-2$s$ and Pb-6$s$, thus enlarge the hybridization between them. This results in the wider O-2$s$ band as also clearly shown in Fig. 1(c).

Similar to above discussions, the up-shifting of Pb-6$p$ bands also decreases the hybridizations between this orbital and O-2$p$/Pb-6$s$ state. Consequently, the Pb-6$s$ and O-2$p$ bands become narrower with higher peak as shown in Fig. 1(d). As will show later, though the hybridization between Pb-6$s$($p$) and O-2$p$ states could still be seen in the figure, the effect on the formation of lone pair is already unfolded.

Above we have demonstrated that our OSEP method does have effect on the hybridizations between atomic states. Then we elaborate the consequence on the Pb-6$s$ lone pair. To do this, we show the electron density contour plot in the $yz$ plane which contains both the Pb and O atom in Fig. 2. As is clear in Fig. 2(a), which illustrates the charge density contour and isosurface of α-PbO from conventional density function calculation without applying any external field, the charge around oxygen atom distribute uniformly like a sphere, whereas an enhanced lobe-like electron density near Pb atom is formed along the $z$ direction and pointing away from PbO layer. Detailed analysis reveals that this lobe-like charge density is formed by the Pb-6$s$ state just below the Fermi level. This is the so called lone pair, as is also shown in previous studies [13, 14].

Since the effect of different field on charge density contour is qualitatively similar, we again show the results with external field to shift the on-site orbital energy by 6 eV in Fig. 2(b)-(d). We first discuss the effects of O-2$s$ and O-2$p$. As mentioned before, O-2$s$ state is quite deep in energy level, therefore it has negligible interaction with other states. Indeed, we find down-shifting O-2$s$ has no effect on the lobe. On the other hand, shifting O-2$p$ has considerable effect on the PDOS of Pb states, as shown in Fig. 1(b). However, as shown in Fig. 2(b), we find that it almost does not affect the size and strength of lobe, indicating O-2$p$ states contribute little to the formation of the lone pair.

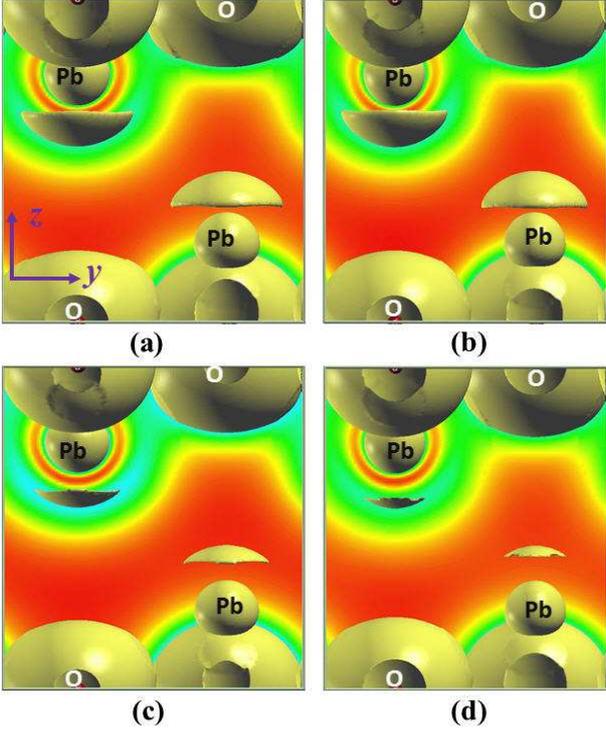

FIG. 2 (color online). Charge density contour and isosurface of α-PbO: (a) without applying field; (b) shifting down O-2$p$; (c) shifting down Pb-6$s$; (d) shifting up Pb-6$p$. The (100) contour plane contains one Pb-O pair (top one), the other Pb-O pair (bottom one) is above the contour plane. Contour levels shown are between 0 (red) and 0.3 e/Å$^3$ (cyan). The lobe-like isosurface (in golden color) near the Pb atom along the $z$ direction indicates the appearance of lone pair.

On the other hand, down-shifting Pb-6$s$ will reduce the size of the lobe [Fig. 2(c)]. Moreover, we find that upshifting Pb-6$p$ also weakens the lone pair as shown in Fig. 2(d). Since O-2$p$ states are not responsible for the lone pair lobe as discussed above, these numerical results clearly demonstrate the importance of Pb-6$s$-6$p$ hybridization on the lone pair.

Following the same strategy, we have also investigated other lone pair active system, such as PbTiO$_3$, Bi$_2$O$_3$ and BiFeO$_3$. We successfully reproduce the asymmetric charge distribution for all of the compounds, and our OSEP calculation confirm that the cation's n$s$ and n$p$ of valence shell participate in forming lobe-like lone pair while oxygen 2$s$ and 2$p$ states only have quite small effect on the lone pair's formation.

As shown in above analysis, we unambiguously determined the importance of cation's outmost n$s$-n$p$ on site hybridizations to the formation of lone pair. Then one question is naturally raised: what is the role of anion, or, why Sn monochalcogenides, SnX (X=O, S, Se and Te), have very different crystalline environment despite having similar electronic configuration? Watson *et al.* compared the electron density of SnX with litharge structure, and found that from SnO to SnTe, a progressively weaker distortion can be seen to form with the Sn electron density becoming more symmetric through the series [20]. They claimed that it is due to the energy levels of anion valence $p$ states increase from O to Te as the group period number getting large, thus the hybridizations between Sn-5$s$ state and X-$p$ states become smaller, consequently, the lone pair becomes weaker from X=O to X=Te [20]. Based on these arguments, they asserted that anion plays an important role in the formation of lone pairs.

To understand the effect of anion, we also performed calculations for SnX. We notice that in the above comparison of the Sn electron density from X=O to X=Te, Watson *et al.* fixed the structures of SnX to be litharge structure and optimized the lattice parameters [20]. The optimized lattice constants and $c/a$ ratio of SnX with litharge structure are significantly different. For example, in our scheme the $c/a$ ratio of optimized SnX structure are 1.234, 0.862, 0.807, 0.734 for X=O, S, Se and Te, respectively. Thus it is important to make clear the role of the structure (volume and $c/a$ ratio) effect on the occurrence of lone pair. We therefore performed two sets of calculation. One is to exactly follow Watson *et al.*'s approach, i.e., using the optimized lattice parameters of SnX in litharge structure. Note that in this way we obtained consistent band structures of SnX with previous work [20]. The calculated charge density contour maps are presented in Fig. 3(a-d). From these figures we can see clearly that as the anion X changes from O to Te, the Sn electron density becomes more symmetric, i.e. the lone pair becomes weaker from SnO to SnTe. These results are in excellent agreement with Watson *et al.*'s results [20].

The other set of calculation is to exclude any structural effects by fixing all the calculations of SnX using the optimized lattice parameters of SnS with litharge structure. The results are shown in Fig. 3(e)-(h). It is interesting to notice that under such circumstance, the lobe becomes larger from SnO to SnTe, which is in the opposite of findings of Watson *et al.* [4, 20] and also our own calculation shown above. Detailed analysis reveals that, from SnO to SnTe, the anion valence $s$ state becomes closer to Sn-5$s$ state, therefore the increased interaction between X valence $s$ state and Sn-5$s$ state pushes up Sn-5$s$ states, making the lone pair more prominent.

To confirm that the anion valence $s$ band indeed has important effect on the formation of lone pair when its energy level is close to that of Sn-5$s$, we again carried out the OSEP calculation by down-shifting X valence $s$ band. As we expect, the down-shift of X valence $s$ band does reduce the lobe, as shown in Fig. S1 of Supporting Material.

Above we have demonstrated that if we fix the structure to be the same for SnX, the lone pair is strengthen instead of being suppressed, then the decrease of lone pair in the first set of calculations, i.e., optimize lattice parameters with litharge structure, can only be attributed to the dramatic decrease of $c/a$ ratio (and $c$ aixs) from SnO to SnTe, as the smaller $c$ apparently will suppress the development of the lone pair.

In order to further check the effect of anion, we have also studied PbO and PbS as another series. Firstly, based on the optimized litharge structure, we reproduce the previous theoretical results [23], i.e. PbO has a strong asymmetric charge distribution around Pb atom, while for PbS, the charge is less asymmetrically distributed as shown in Fig. S2(a) and S2(b) of Supporting Material. Same as the case of SnX, the optimized litharge structures are dramatically different ($a=3.95$, $c=4.87$ Å for PbO, and $a=4.99$,

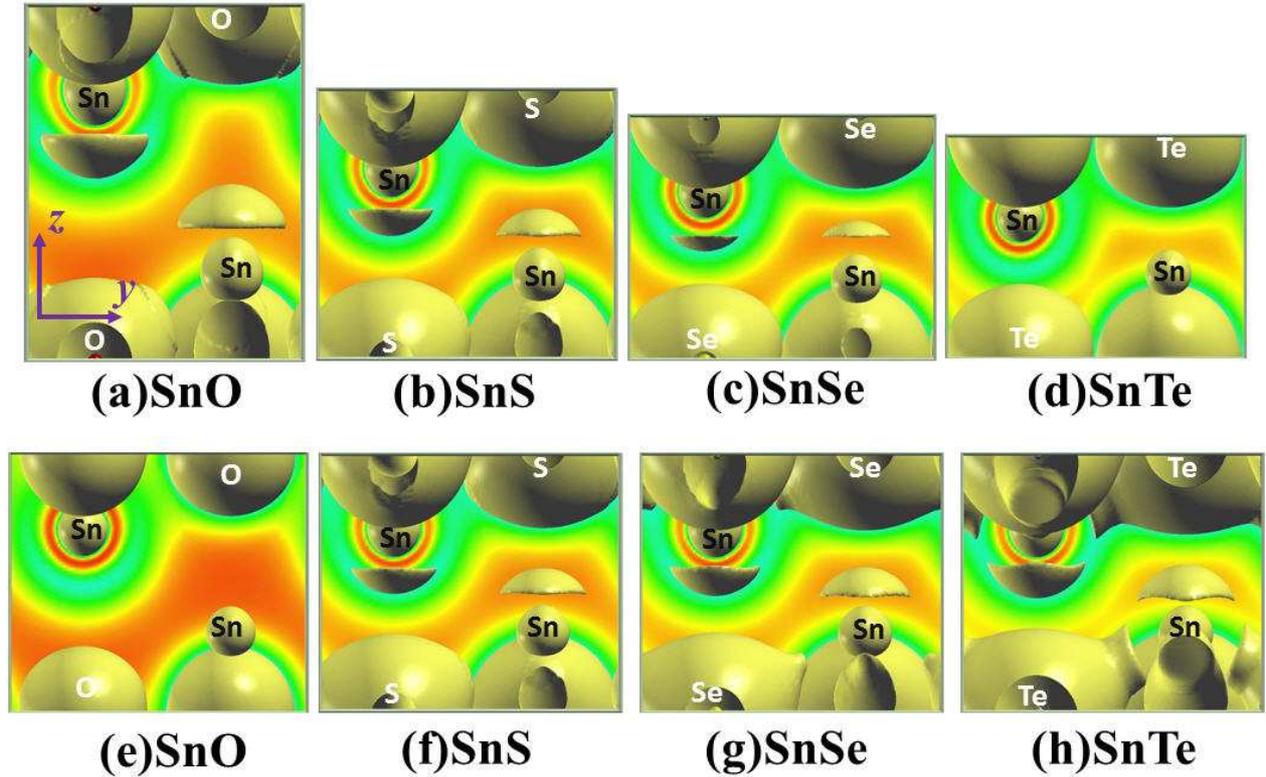

FIG. 3 (color online). Charge density contour and isosurface of SnX (X=O,S,Se,Te). (a)SnO, (b)SnS, (c)SnSe and (d)SnTe represent SnX in their own optimized litharge structure, and (e)SnO, (f)SnS, (g)SnSe and (h)SnTe are obtained using the exactly same structure of optimized SnS's litharge structure. Contour levels shown in the (100) contour plane are between 0 (red) and 0.3 e/Å$^3$ (cyan).

$c=3.85$ Å for PbS). Thus, we also perform calculation for PbO using the optimized structure of PbS with litharge structure, the numerical calculation confirm that similar to the case of SnX, this can also be attributed to structural effect, and the lone pair is now found to be less obvious than that of PbS, as clearly shown in Fig. S2(c) and S2(d) of Supporting Material.

Finally, we would like to explain the unusual experimentally observed structural transitions in lone pair systems like Sn monochalcogenides. As mentioned before, using the OSEP method we can control the formation and strength of lone pair. Then we have performed two sets of OSEP calculations on SnO, one is for litharge structure and the other is for rocksalt structure with inversion center. We continuously applied OSEP to shift down the Sn-5$s$ state in order to suppress the lone pair. As clearly shown in Fig. 4, with the increase of external potential, the energy difference between the SnO with litharge structure and rocksalt structure gradually decreases. At around 15 V applied external potential (namely, we apply external field to downshift the Sn-5$s$ state by 15 eV), where the lone pair disappears, the system undergoes a structural transformation from litharge structure to rocksalt structure. This clearly reveals the relationship between the structural instabilities and the lone pair in Sn monochalcogenides.

We also performed similar calculations by shifting the Sn-5$p$ and O-2$p$. It is found that the effect of Sn-5$p$ is similar to the Sn-5$s$, whereas shifting O-2$p$ does not lead to the structural transition. These results again verify our above results that the O-2$p$ does not participate in the formation of lone pair.

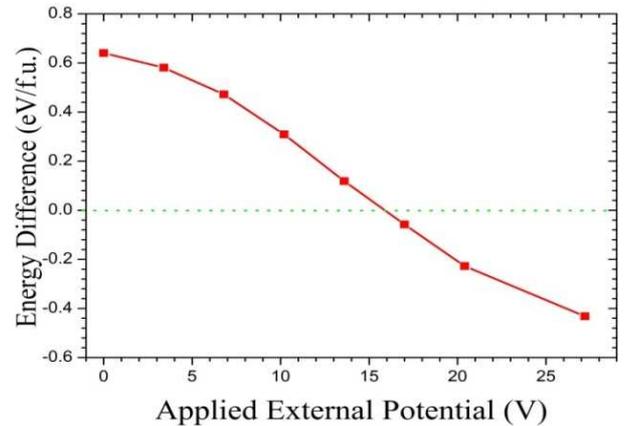

FIG. 4 (color online). Energy Difference between the SnO with litharge structure and rocksalt structure with continuous applied external potential to shift-down Sn-5$s$ state. At around 15 V, where the lone pair is sufficiently suppressed, SnO transforms from litharge structure to rocksalt structure with inversion center.

Using orbital selective external potential method, we have investigated the origin of stereochemically active lone pair. After careful studies on the electronic structures of α-PbO and SnX (X=O, S, Se, Te), we find that the response of cation's n$s$ and n$p$ subshell are the origin of the formation of lone pair, while the anion's n$s$ and n$p$ orbitals have much less effects on properties of lone pair. In short, lone pair is formed largely due to the hybridization of cati-on's n$s$ and n$p$ valence shell. In addition, we also find that the lone pair is responsible for the structural transition in Sn and Pb monochalcogenides.

Yongping Du and Hang-Chen Ding contributed equally to this work. The work was supported by the National Key Project for Basic Research of China (Grants No. 2011CB922101, No. 2010CB923404 and 2013CB922300), NSFC under Grants No. 91122035, 11174124, 10974082, 61125403, and PCSIRT, NCET, PAPD project. Computations were performed at the ECNU computing center. H.-C. Ding acknowledges the support from ECNU-PY2012001.

*Corresponding author.
xgwan@nju.edu.cn
wxbdcg@gmail.com or cgduan@clpm.ecnu.edu.cn